\documentclass[12pt,a4paper]{article}
\usepackage{fullpage}

\usepackage{amsmath}
\usepackage{mathtools}
\usepackage{amsthm}
\usepackage{amssymb}
\usepackage{amsfonts}
\usepackage{mleftright}
\usepackage{enumitem}  
\usepackage{graphicx}
\usepackage{systeme}
\usepackage{ltablex,booktabs,ragged2e,caption}

\usepackage[table]{xcolor}
\theoremstyle{definition}
\newtheorem{example}{Example}

\makeatletter
\newcommand{\longdash}[1][2em]{%
	\makebox[#1]{$\m@th\smash-\mkern-7mu\cleaders\hbox{$\mkern-2mu\smash-\mkern-2mu$}\hfill\mkern-7mu\smash-$}}

\makeatletter
\renewcommand*\env@matrix[1][\arraystretch]{%
	\edef\arraystretch{#1}%
	\hskip -\arraycolsep
	\let\@ifnextchar\new@ifnextchar
	\array{*\c@MaxMatrixCols c}}
\makeatother

\begin{document}

\title{Multi-Message Private Information Retrieval using Product-Matrix {MSR} and {MBR} Codes}

\author{Chatdanai Dorkson\footnote{Department of Mathematics, Royal Holloway, University of London, Egham, Surrey, TW20 0EX, United Kingdom. Chatdanai.Dorkson.2016@rhul.ac.uk. This author is supported by the Development and Promotion of Science and Technology Talents Project (Royal Government of Thailand scholarship).}, Siaw-Lynn Ng\footnote{Information Security Group, Royal Holloway, University of London, Egham, Surrey, TW20 0EX, United Kingdom. S.Ng@rhul.ac.uk.}}
\date{\today}
\maketitle

\begin{abstract}
	Multi-message private information retrieval (MPIR) is an interesting variation of PIR which allows a user to download multiple messages from the database without revealing the identity of the desired messages. Obviously, the user can repeatly use a single-message PIR scheme, but we wish for a more efficient way to reduce the download cost. In \cite{multi}, Banawan and Ukulus investigate the multi-message PIR problem with replicated database. In this paper, we consider multi-message PIR schemes where the database is stored using minimum storage regenerating (MSR) or minimum bandwidth regenerating (MBR) codes which are classes of optimal regenerating codes providing efficient repair when a node in the system fails. The relationships between the costs of storage, retrieval and repair are analysed, and explicit schemes using the MSR and MBR codes from \cite{matrix}, which both achieve the optimal curve of the trade-off, are given. To the best of our knowledge, our work is the first to explore the multi-message PIR with coded database.
\end{abstract}


\section{Introduction}
\label{sec1}

Private information retrieval (PIR) schemes allow a user to download files from the database without revealing any information on which records a user wants to retrieve. In the original setting for PIR \cite{Original}, the whole database is replicated among $n$ non-colluding nodes, which results in high storage cost, so this motivates the use of erasure codes which means that only a fraction of the entire database is stored in each node, and this is called \textit{code-based PIR schemes}. 

In \cite{extra}, Shah et al. present the first work of the code-based PIR schemes proving that only an extra bit of download is needed to retrieve the desired record, and they also provide another PIR scheme using the product-matrix minimum bandwidth regenerating (MBR) codes \cite{matrix}. Chan et al. \cite{tho} give retrieval schemes for a general class of linear storage codes, and discover the relationship between storage and retrieval cost in the context of their proposed PIR schemes, and subsequently, Tajeddine and Rouayheb \cite{MDS} design an explicit scheme using MDS codes achieving the optimal curve of the trade-off in \cite{tho}. Later, Kumar et al. \cite{arbitrary} propose PIR schemes that use an arbitrary systematic linear storage code of rate $ > 1/2$, and, interestingly, locally repairable codes (LRCs) \cite{LRCs} and Pyramid codes \cite{pyramid}, which have more efficient repair property, can be used to achieve the optimal scheme.

As the classical PIR setting has been extended to many variations, one interesting scenario is when a user wants to retrieve more than one record. Clearly, the user can use a single-message scheme multiple times, but is there a more efficient way to do this? This is the multi-message PIR (MPIR) problem. In \cite{multi}, Banawan and Ulukus consider the problem of capacity which is defined as the maximum of the retrieval rate over all possible PIR schemes by analysing the capacity of multi-message PIR schemes with replicated database, and give a capacity-achievable scheme when the number of desired records is at least half of the number of total records.

In this paper, we propose the general multi-message PIR model where the product-matrix regenerating codes is used for storage. The use of regenerating codes beneficially reduces the repair cost when a node failure occurs in the system, hence our scheme obtains more efficient repair compared to schemes using MDS codes. To the best of our knowledge, \cite{extra} and \cite{MSR} are the only paper that uses regenerating codes in their PIR scheme, and our work is the first to explore multi-message PIR with coded databases. Furthermore, we analyse the relationship between the costs of storage, retrieval and repair, and design explicit schemes that fit the optimal curve of the trade-off using the product-matrix MSR and MBR codes from \cite{matrix}.

The organisation of this paper is as follows. We recall the MSR and MBR codes, and the product-matrix constructions from \cite{matrix} in Section 2. The system model of multi-message PIR scheme using product-matrix regenerating codes is then given in Section 3. In Section 4, we obtain the decodability condition and trade-off analysis between storage, retrieval, and repair costs in the system. Motivating examples and explicit constructions of our optimal MPIR schemes using MSR and MBR codes are presented in Section 5. We give the discussion on our constructions in Section 6. Lastly, in the Appendix A, we propose an alternative optimal MPIR scheme using product-matrix MSR codes with a different retrieval pattern. This scheme has slightly higher cPoP and lower storage overhead compared to the scheme using MSR codes in Section 5, and it turns out to be a generalisation of our single-message construction in \cite{MSR}.

\section{Product-Matrix {MSR} and {MBR} Codes}

\subsection{{MSR} and {MBR} Codes}

An $(n,k,r,\alpha,\beta,B)$ \textit{regenerating code} \cite{dimakis} is defined to be a distributed storage code storing the database of size $B$ among $n$ nodes where each node stores $\alpha$ symbols satisfying two properties: (i)(recovery) The entire database can be recovered from the data stored in any $k$ nodes; (ii)(repair) If one of the storage nodes fails, then a newcomer node connects to some set of $r$ remaining nodes where $k<r<n$, and downloads $\beta$ symbols from each of these $r$ nodes in order to regenerate $\alpha$ symbols in such a way that we can perform (i) and (ii) again when another node failure occurs.

The total amount of $r\beta$ symbols downloaded for regenerating is called the \textit{repair bandwidth}, and typically the repair bandwidth is smaller than the size of the whole database. There are various repair models, but for PIR we focus on the exact repair model, where a newcomer node will regenerate the same data as was stored in the failed node in order to maintain the initial state of the storage nodes.

In \cite{YWu}, the parameters of a regenerating code is shown to necessarily satisfy $$ B \leq \sum_{i=0}^{k-1} \min\{\alpha,(r-i)\beta\},$$ 
and the achievable trade-off between storage overhead and repair bandwidth is characterised by fixing the repair bandwidth, and then deriving the minimum $\alpha$ which satisfies the above equation. Two interesting extremal points on the optimal trade-off curve are the \textit{minimum storage regeneration} (MSR) point which minimises storage overhead first and then minimises repair bandwidth, and the \textit{minimum bandwidth regeneration} (MSR) point which minimises in the reverse order. It can be shown that the MSR point is achieved by $$(\alpha_{MSR},\beta_{MSR}) = \bigg(\frac{B}{k}, \frac{B}{k(r-k+1)}\bigg),$$ and \textit{MSR codes} are $(n,k,r,\alpha,\beta,B)$ regenerating codes that satisfies the above equation. Also the MBR point is achieved by $$(\alpha_{MBR},\beta_{MBR}) = \bigg(\frac{2rB}{k(2r-k+1)}, \frac{2B}{k(2r-k+1)}\bigg),$$ and \textit{MBR codes} are $(n,k,r,\alpha,\beta,B)$ regenerating codes that satisfies the above equation.

\subsection{The Product-Matrix {MSR} Codes (\cite{matrix})}

Under the product-matrix framework, each codeword is represented by an $(n \times \alpha)$ \textit{code matrix} $C$ which is the product $$C = \Psi \cdot M$$ of an $(n \times r)$ \textit{encoding matrix} $\Psi$ and an $(r \times \alpha)$ \textit{message matrix} $M$. The message matrix $M$ contains the $B$ message symbols. In the code matrix $C$, row $i$ consists of the $\alpha$ encoded symbols stored by node $i$ for each $i \in [n]$.

In \cite{matrix}, Rashmi, Shah and Kumar gave an explicit construction for the MSR code with $r=2k-2$, so the parameters are $(n,k,r,\alpha,\beta,B) = (n,k,2k-2,k-1,1,k(k-1))$ where $n>2k-2$ using the product-matrix framework. First, they let the encoding matrix $\Psi$ be any $(n \times r)$ matrix given by
$$ \Psi = \begin{bmatrix} 
\Phi & \Lambda\Phi \\
\end{bmatrix}$$
where $\Phi$ is an $(n \times \alpha)$ matrix and $\Lambda$ is an $(n \times n)$ diagonal matrix such that (i) any $r$ rows of $\Psi$ are linearly independent, (ii) any $k-1$ rows of $\Phi$ are linearly independent, (iii) the $n$ diagonal elements of $\Lambda$ are all distinct. The rows of $\Psi$ are denoted by $\Psi_i, i \in [n]$.
Next, the $(r \times \alpha)$ message matrix $M$ is defined as 
$$ M = \begin{bmatrix} 
S_1 \\
S_2
\end{bmatrix}$$
where $S_1$ and $S_2$ are $(\alpha \times \alpha)$ symmetric matrices constructed such that ${k \choose 2}$ entries in the upper-triangular part of each matrix are filled up by ${k \choose 2}$ distinct message symbols and entries in the strictly lower-triangular are chosen to make the matrices symmetric. This is the MSR code we will use in Section 5.1.

\subsection{The Product-Matrix {MBR} Codes (\cite{matrix})}

Rashmi, Shah and Kumar also gave an explicit construction for the MBR code with parameters $$(n,k,r,\alpha,\beta,B) = \bigg( n,k,r,r,1,\frac{k(2r-k+1)}{2}\bigg)$$ where $n>r$ using the product-matrix framework in \cite{matrix}. First, the encoding matrix $\Psi$ is an $(n \times r)$ matrix given by
$$ \Psi = \begin{bmatrix} 
\Phi & \Delta \\
\end{bmatrix}$$
where $\Phi$ is an $(n \times k)$ matrix and $\Delta$ is an $(n \times (r-k))$ matrix such that (i) any $r$ rows of $\Psi$ are linearly independent, (ii) any $k$ rows of $\Phi$ are linearly independent. The rows of $\Psi$ are denoted by $\Psi_i, i \in [n]$.
Next, the $(r \times r)$ message matrix $M$ is defined as 
$$ M = \begin{bmatrix} 
S_1 & S_2 \\
S_2^T & 0
\end{bmatrix}$$
where $S_1$ is a $(k \times k)$ matrix constructed such that ${k+1 \choose 2}$ entries in the upper-triangular part of each matrix are filled up by ${k+1 \choose 2}$ distinct message symbols and entries in the strictly lower-triangular are chosen to make the matrices symmetric, and the $(k \times (r-k))$ matrix $S_2$ are filled up by the remaining $k(r-k)$ message symbols. This is the MBR code we will use in Section 5.2.

\section{System Model}

In this section, we formally present the storage model and its retrieval scheme. Consider there are $n$ non-communicating nodes in the system that store a database $X$ which consists of $m$ records, each of length $\ell$, denoted by $X^1, X^2, \dots, X^m \in \mathbb{F}_q^\ell$. Each record is encoded and distributed across $n$ nodes by the same product-matrix regenerating code with parameters $(n, k , r, \alpha, \beta, \ell)$ which can be written as $$C^j = \Psi \cdot \mathcal{M}^j$$ where $\mathcal{M}^j$ is the corresponding message matrix of $X^j$. Write $$\mathcal{M} = \begin{bmatrix} 
\mathcal{M}^1 & \cdots & \mathcal{M}^m \\
\end{bmatrix},$$ and denote by $\mathcal{M}_i$ the row $i$ of $\mathcal{M}$. Hence, we can see the entire system as $$C= \begin{bmatrix} 
C^1 & \cdots & C^m \\
\end{bmatrix} = \begin{bmatrix} 
\Psi \cdot \mathcal{M}^1 & \cdots & \Psi \cdot \mathcal{M}^m \\
\end{bmatrix} = \Psi \cdot \mathcal{M},$$ and each node stores $m\alpha$ symbols in total. We denote by $C_i$ the row $i$ of $C$ which is all symbols stored in node $i$, and $C_i^j$ the row $i$ of $C^j$ which is all symbols of $X^j$ stored in node $i$.

We assume that in the retrieval step the user wants to download $p$ records when $p \leq m$, denoted by $X^{f_1}, \dots, X^{f_p}$. The user submits a $d \times m\alpha$ query matrix $Q^i$ over $GF(q)$ to node $i$. We can interpret $d$ rows of $Q^i$ as $d$ subqueries, and for instance $d$ is set to be $\alpha$ in our constructions. Finally, node $i$ computes and responds with an answer $A_i^T = Q^i C_i^T$. The retrieval steps are as follows:

\begin{enumerate}[label=(\roman*)]
	\item (Initialisation) The user generates an $d \times m\alpha$ matrix $U$ whose elements are chosen independently and uniformly at random over $GF(q)$. Let $U_j$ be row $j$ of $U$.
	\item (Query Generation) The query matrix $Q^h$ is defined by $d \times \alpha$ binary matrices $$V^{h_{f_1}} , \dots, V^{h_{f_p}},$$ as $$Q^h = U + V^{h_{f_1}} E^{f_1} + \dots + V^{h_{f_p}} E^{f_p}$$ where $$E^{f_i} = \left[\begin{array}{c|c|c}
	\textbf{0}_{\alpha \times (f_i-1)\alpha} & I_{\alpha \times \alpha} & \textbf{0}_{\alpha \times (m-f_i)\alpha}
	\end{array}\right].$$ In other words, $E^{f_i}$ is an $\alpha \times m\alpha$ matrix such that $C_h (E^{f_i})^T = C_h^{f_i}$ which is a coded data piece of a desired record $X^{f_i}$ stored in node $h$. If the entry $(a,b)$ of $V^{h_{f_i}}$ is 1, then it implies that the entry $C^{f_i}_{hb}$ is privately retrieved by the $a^{th}$ subquery of $Q^h$.
	\item (Response Mappings) Each node $h$ returns $A_h^T = Q^h C_h^T$.
\end{enumerate}
Let $H(\cdot)$ be the entropy function. An MPIR scheme is said to be a \textit{perfect information-theoretic PIR scheme} if 
\begin{enumerate}
	\item
	(i)(privacy) $H(f_1,\dots,f_p|Q_i)=H(f_1,\dots,f_p)$ for every $i \in [n]$;
	
	\item
	(ii)(decodability) $H(X^{f_1},\dots,X^{f_p}|A_1,\dots,A_n)=0$.
	
\end{enumerate}
According to our definition, (i) implies that a node $i$ does not obtain any information about which records are being downloaded by the user, and (ii) ensures that the user can recover the desired records $X^{f_1},\dots,X^{f_p}$ with no errors from all responses $A_i, i \in [n]$.
\vspace{-10pt}\\

To measure the efficiency of the MPIR scheme, we use three metrics, namely \textit{Storage Overhead (SO)}, \textit{communication Price of Privacy (cPoP)} and \textit{Repair Ratio (RR)}. First, SO is defined to be the ratio of the total storage used in the scheme to the total size of the whole database which is $$SO = n(m\alpha) / m\ell = n\alpha/\ell$$ in our model, and the cPoP is defined in \cite{MDS} as the ratio of the total amount of downloaded data to the total size of all desired records which, in our model, is $$cPoP = dn/p\ell.$$ Lastly, RR is defined in our paper \cite{MSR} as the ratio of the total amount of symbols downloaded for repairing a failed node to the size of the failed node which is equal to $$RR = mr/m\alpha = r/\alpha$$ in our model.

\section{Decodability Condition and Trade-off Analysis}
From the retrieval scheme, we can see that in fact, the response from node $i$ is
\begin{align*}
A_i &= C_i(Q^i)^T \\
&= C_i [U^T+(E^{f_1})^T(V^{i_{f_1}})^T + \dots +(E^{f_p})^T (V^{i_{f_p}})^T] \\
&= C_i U^T + C_i (E^{f_1})^T (V^{i_{f_1}})^T + \dots +  C_i (E^{f_p})^T (V^{i_{f_p}})^T  \\
&= C_i U^T + (C_i^{f_1})(V^{i_{f_1}})^T + \dots + (C_i^{f_p})(V^{i_{f_p}})^T.
\end{align*}
Then, the $j^{th}$ response in $A_i$ is $A_{ij} = C_i (U_j)^T + (C_i^{f_1})(V_j^{i_{f_1}})^T + \dots + (C_i^{f_p})(V_j^{i_{f_p}})^T$ where $V^{i_{f_u}}_j$ is the row $j$ of $V^{i_{f_u}}, u \in [p]$. Hence, records $X^{f_1}, \dots, X^{f_p}$ should be decoded by solving the system of linear equations $$A_{ij} = C_i (U_j)^T + (C_i^{f_1})(V_j^{i_{f_1}})^T + \dots + (C_i^{f_p})(V_j^{i_{f_p}})^T,$$ for all $i \in [n], j \in [d]$ where the unknowns are $$(C_i (U_j)^T, i \in [n], j \in [d], C^{f_u}_{ab}, u \in [p], a \in [n], b \in [\alpha]).$$ Consider first the unknowns $C_i (U_j)^T, i \in [n], j \in [d]$, we can see that for each $j \in [d],$
\begin{align*}
C(U_j)^T &= \Psi \cdot \mathcal{M} \cdot (U_j)^T \\
&= \Psi \cdot \begin{bmatrix} 
I_1^j & \cdots & I_r^j \end{bmatrix}^T
\end{align*}
where $I_h^j = \mathcal{M}_h \cdot (U_j)^T, h \in [r].$
For the unknowns $C^{f_u}_{ab}, u \in [p], a \in [n], b \in [\alpha]$, we know that $$C^{f_u}_{ab} = \mbox{the entry }(a,b) \mbox{ of } \Psi \cdot \mathcal{M}^{f_u},\mbox{ }\forall u \in [p], a \in [n], b \in [\alpha].$$ 
Hence, the retrieval scheme is decodable if the following system of linear equations
\[
\left\{
\begin{aligned} 
A_{ij} &= C_i (U_j)^T + (C_i^{f_1})(V_j^{i_{f_1}})^T + \dots + (C_i^{f_p})(V_j^{i_{f_p}})^T, && \forall i \in [n], j \in [d] \\ 
C_s (U_t)^T &= \Psi_s \cdot \begin{bmatrix}
 I_1^t & \cdots & I_r^t \end{bmatrix}^T, && \forall s \in [n], t \in [d] \\ 
C^{f_u}_{ab} &= \mbox{the entry }(a,b) \mbox{ of } C^{f_u}, && \forall u \in [p], a \in [n], b \in [\alpha]  \end{aligned} 
\right.
\] has a unique solution, where the unknowns are $$( C_i (U_j)^T, i \in [n], j \in [d], C^{f_u}_{ab}, u \in [p], a \in [n], b \in [\alpha]).$$ This condition is called \textit{decodability condition}.
\\
\vspace{-5pt}

Next, we will give the trade-off analysis between storage overhead and cPoP. First, we count the number of unknowns in the system of linear equations in the decodability condition which is equal to $nd+pn\alpha$. Next, we count the number of linearly independent equations in the system. Consider $$C_s (U_t)^T = \Psi_s \cdot \begin{bmatrix} I_1^t & \cdots & I_r^t \end{bmatrix}^T, \forall s \in [n], t \in [d],$$ so we have, for each $t \in [d]$, $$C(U_t)^T = \Psi \cdot \begin{bmatrix} 
I_1^t & \cdots & I_r^t \end{bmatrix}^T.$$ Since $\Psi$ is of rank $r$, it has a parity check matrix $P$ of rank $n-r$ such that $P \cdot \Psi = 0$. So we have $$P \cdot C(U_t)^T = P \cdot \Psi \cdot \begin{bmatrix} 
I_1^t & \cdots & I_r^t \end{bmatrix}^T = 0.$$ This gives us $n-r$ linearly independent equations for each $t \in [d]$. Then, for $$C^{f_u}_{ab} = \mbox{the entry }(a,b) \mbox{ of } C^{f_u}, \forall u \in [p], a \in [n], b \in [\alpha],$$ since any $k$ rows of $C^{f_u}$ would give us $\mathcal{M}^{f_u}$, the remaining $n-k$ rows must be able to be written in terms of linear combinations of those $k$ rows of $C^{f_u}$. This give us $(n-k)\alpha$ equations in $C^{f_u}_{ab}$. Hence, there are at most $nd+(n-r)d+p(n-k)\alpha$ linearly independent equations in the system. If the retrieval scheme meets the decodability condition, then $$nd+pn\alpha \leq nd+(n-r)d+p(n-k)\alpha,$$ which implies that $$pk\alpha \leq (n-r)d.$$ Therefore, $$1 \leq \frac{dn}{p\ell} \cdot \frac{\ell}{n\alpha} \cdot \frac{n-r}{k}.$$ \\
In terms of storage overhead and cPoP we have $$1 \leq cPoP  \bigg( \frac{n-r}{k (SO)} \bigg).$$ This shows that there is a trade-off between cPoP and storage overhead, and in terms of repair ratio and cPoP we have $$1 \leq cPoP\bigg(\frac{\ell}{k\alpha}\bigg) - RR\bigg(\frac{d}{pk}\bigg).$$ This shows that cPoP is bounded below by repair ratio.

\section{Our constructions}

\subsection{An MPIR scheme using a product-matrix MSR code}

In this construction, we use the product-matrix MSR code from \cite{matrix} with $$n=pk+r=pk+(2k-2),$$ over the finite field $\mathbb{F}_q$, so the parameters of the MSR code are $$(n,k,r,\alpha,\beta,\ell)=(pk+(2k-2),k,2k-2,k-1,1,k(k-1)).$$ We first start with an example to motivate our scheme.

\begin{example}
Suppose that we have 3 records over the finite field $\mathbb{F}_{13}$, each with size $6$, which can be written as $$X^i = \{x_{i1}, x_{i2}, x_{i3}, x_{i4}, x_{i5}, x_{i6}\}, \mbox{ for } i=1,2,3.$$
We use a $(10, 3, 4, 2, 1, 6)$ product-matrix MSR code over $\mathbb{F}_{13}$ to encode each record by choosing the encoding matrix $\Psi$ to be the Vandermonde matrix, and the message matrix $\mathcal{M}^i$ for the record $i , i \in \{1,2,3\}$ as described in Section 2.2:
\[ \Psi =
\begin{bmatrix}
1 & 1 & 1 & 1 & 1 & 1 & 1 & 1 & 1 & 1\\
1 & 2 & 3 & 4 & 5 & 6 & 7 & 8 & 9 & 10\\
1 & 4 & 9 & 3 & 12 & 10 & 10 & 12 & 3 & 9\\
1 & 8 & 1 & 12 & 8 & 8 & 5 & 5 & 1 & 12
\end{bmatrix}^T, \quad  \mathcal{M}^i = \begin{bmatrix}
x_{i1} & x_{i2} \\
x_{i2} & x_{i3} \\
x_{i4} & x_{i5} \\
x_{i5} & x_{i6} 
\end{bmatrix}.
\]
Hence, each node stores
\begin{center}
	\scalebox{0.6}{
		\begin{tabular}{ |c|c|c|c|c| } 
			\hline
			node 1 & node 2 & node 3 & node 4 & node 5 \\
			\hline
			$x_{11}+x_{12}+x_{14}+x_{15}$ & $x_{11}+2x_{12}+4x_{14}+8x_{15}$ & $x_{11}+3x_{12}+9x_{14}+x_{15}$ & $x_{11}+4x_{12}+3x_{14}+12x_{15}$ & $x_{11}+5x_{12}+12x_{14}+8x_{15}$ \\ 
			$x_{12}+x_{13}+x_{15}+x_{16}$ & $x_{12}+2x_{13}+4x_{15}+8x_{16}$ & $x_{12}+3x_{13}+9x_{15}+x_{16}$ & $x_{12}+4x_{13}+3x_{15}+12x_{16}$ & $x_{12}+5x_{13}+12x_{15}+8x_{16}$ \\ 
			\hline
			$x_{21}+x_{22}+x_{24}+x_{25}$ & $x_{21}+2x_{22}+4x_{24}+8x_{25}$ & $x_{21}+3x_{22}+9x_{24}+x_{25}$ & $x_{21}+4x_{22}+3x_{24}+12x_{25}$ & $x_{21}+5x_{22}+12x_{24}+8x_{25}$ \\ 
			$x_{22}+x_{23}+x_{25}+x_{26}$ & $x_{22}+2x_{23}+4x_{25}+8x_{26}$ & $x_{22}+3x_{23}+9x_{25}+x_{26}$ & $x_{22}+4x_{23}+3x_{25}+12x_{26}$ & $x_{22}+5x_{23}+12x_{25}+8x_{26}$ \\ 
			\hline
			$x_{31}+x_{32}+x_{34}+x_{35}$ & $x_{31}+2x_{32}+4x_{34}+8x_{35}$ & $x_{31}+3x_{32}+9x_{34}+x_{35}$ & $x_{31}+4x_{32}+3x_{34}+12x_{35}$ & $x_{31}+5x_{32}+12x_{34}+8x_{35}$ \\ 
			$x_{32}+x_{33}+x_{35}+x_{36}$ & $x_{32}+2x_{33}+4x_{35}+8x_{36}$ & $x_{32}+3x_{33}+9x_{35}+x_{36}$ & $x_{32}+4x_{33}+3x_{35}+12x_{36}$ & $x_{32}+5x_{33}+12x_{35}+8x_{36}$ \\ 
			\hline
		\end{tabular}}
	\end{center}
	
\begin{center}
	\scalebox{0.6}{
		\begin{tabular}{ |c|c|c|c|c| } 
			\hline
			node 6 & node 7 & node 8 & node 9 & node 10 \\
			\hline
			  $x_{11}+6x_{12}+10x_{14}+8x_{15}$ & $x_{11}+7x_{12}+10x_{14}+5x_{15}$ & $x_{11}+8x_{12}+12x_{14}+5x_{15}$ & $x_{11}+9x_{12}+3x_{14}+x_{15}$ & $x_{11}+10x_{12}+9x_{14}+12x_{15}$ \\ 
			$x_{12}+6x_{13}+10x_{15}+8x_{16}$ & $x_{12}+7x_{13}+10x_{15}+5x_{16}$ & $x_{12}+8x_{13}+12x_{15}+5x_{16}$ & $x_{12}+9x_{13}+3x_{15}+x_{16}$ & $x_{12}+10x_{13}+9x_{15}+12x_{16}$ \\ 
			\hline
		     $x_{21}+6x_{22}+10x_{24}+8x_{25}$ & $x_{21}+7x_{22}+10x_{24}+5x_{25}$ & $x_{21}+8x_{22}+12x_{24}+5x_{25}$ & $x_{21}+9x_{22}+3x_{24}+x_{25}$ & $x_{21}+10x_{22}+9x_{24}+12x_{25}$ \\ 
			 $x_{22}+6x_{23}+10x_{25}+8x_{26}$ & $x_{22}+7x_{23}+10x_{25}+5x_{26}$ & $x_{22}+8x_{23}+12x_{25}+5x_{26}$ & $x_{22}+9x_{23}+3x_{25}+x_{26}$ & $x_{22}+10x_{23}+9x_{25}+12x_{26}$ \\ 
			\hline
			 $x_{31}+6x_{32}+10x_{34}+8x_{35}$ & $x_{31}+7x_{32}+10x_{34}+5x_{35}$ & $x_{31}+8x_{32}+12x_{34}+5x_{35}$ & $x_{31}+9x_{32}+3x_{34}+x_{35}$ & $x_{31}+10x_{32}+9x_{34}+12x_{35}$ \\ 
			 $x_{32}+6x_{33}+10x_{35}+8x_{36}$ & $x_{32}+7x_{33}+10x_{35}+5x_{36}$ & $x_{32}+8x_{33}+12x_{35}+5x_{36}$ & $x_{32}+9x_{33}+3x_{35}+x_{36}$ & $x_{32}+10x_{33}+9x_{35}+12x_{36}$ \\ 
			\hline
		\end{tabular}}
	\end{center} 
	Recall that $C^a_{ij}$ is the $j^{th}$ symbol of record $a$, stored in node $i$. Here the entire database can be recovered from the content of any 3 nodes, and if any one node failed, it can be repaired by downloading one symbol each from 4 of the remaining nodes.
\\
\vspace{-5pt}

	In the retrieval step, suppose the user wants record $X^1$ and $X^2$. The query $Q^i$ is a $(2 \times 6)$ matrix which we can interpret as $2$ subqueries submitted to node $i$ for each $i \in [10]$. To form the query matrices, the user generates a $(2 \times 6)$ random matrix $U = [u_{ij}]$ whose entries are chosen uniformly at random from $\mathbb{F}_{13}$. Recall that $V^{i_j}$ is a matrix which is part of the query submitted to node $i$, attempting to retrieve information about record $X^j$. Choose  \[
	V^{1_1} =  V^{2_1} = V^{3_1} = V^{4_2} = V^{5_2} = V^{6_2} = \begin{bmatrix}
	1 & 0 \\
	0 & 1
	\end{bmatrix},\] and
	$$V^{1_2} = V^{2_2} = V^{3_2} = V^{4_1} = V^{5_1} = V^{6_1} = $$
	$$V^{7_1} = V^{7_2} = V^{8_1} = V^{8_2} = V^{9_1} = V^{9_2} = V^{10_1} = V^{10_2} = \textbf{0}_{2 \times 2}.$$
	As \[
	E^1 = \begin{bmatrix}
	1 & 0 & 0 & 0 & 0 & 0 \\
	0 & 1 & 0 & 0 & 0 & 0 \\  
	\end{bmatrix}, \quad E^2 = \begin{bmatrix}
	0 & 0 & 1 & 0 & 0 & 0 \\
	0 & 0 & 0 & 1 & 0 & 0 \\  
	\end{bmatrix},
	\]
	we have
	\[
	V^{1_1}E^1 = V^{2_1}E^1 = V^{3_1}E^1 = \begin{bmatrix}
	1 & 0 & 0 & 0 & 0 & 0 \\
	0 & 1 & 0 & 0 & 0 & 0 
	\end{bmatrix}
	\]
		\[
		V^{4_2}E^2 = V^{5_2}E^2 = V^{6_2}E^2 = \begin{bmatrix}
		0 & 0 & 1 & 0 & 0 & 0 \\
		0 & 0 & 0 & 1 & 0 & 0
		\end{bmatrix}
		\]
	and $$V^{1_2}E^2 = V^{2_2}E^2 = V^{3_2}E^2 = V^{4_1}E^1 = V^{5_1}E^1 = V^{6_1}E^1 = $$
	$$V^{7_1}E^1 = V^{7_2}E^2 = V^{8_1}E^1 = V^{8_2}E^2 = V^{9_1}E^1 = V^{9_2}E^2 = V^{10_1}E^1 = V^{10_2}E^2 = \textbf{0}_{2 \times 6}.$$

	\noindent The query matrices are $Q^i = U+V^{i_1}E^1+V^{i_2}E^2 , i \in [10]$. Then each node computes and returns the length-$2$ vector $A_i^T = Q^i C^T_i$. Write $A_i = (A_{i1},A_{i2})$. Recall that 
	\begin{align*}
		\mathcal{M}_1 &= (x_{11}, x_{12}, x_{21}, x_{22}, x_{31}, x_{32}), \\
		\mathcal{M}_2 &= (x_{12}, x_{13}, x_{22}, x_{23}, x_{32}, x_{33}), \\
		\mathcal{M}_3 &= (x_{14}, x_{15}, x_{24}, x_{25}, x_{34}, x_{35}), \\
		\mathcal{M}_4 &= (x_{15}, x_{16}, x_{25}, x_{26}, x_{35}, x_{36}).
	\end{align*}
	Consider first subquery 1, we obtain
	\begin{align*}
		C^1_{11}+I^1_1+I^1_2+I^1_3+I^1_4 &= A_{1,1},  \tag{1}\\
		C^1_{21}+I^1_1+2I^1_2+4I^1_3+8I^1_4 &= A_{2,1},  \tag{2}\\
		C^1_{31}+I^1_1+3I^1_2+9I^1_3+I^1_4 &= A_{3,1},  \tag{3}\\
		C^2_{43}+I^1_1+4I^1_2+3I^1_3+12I^1_4 &= A_{4,1},  \tag{4}\\
		C^2_{53}+I^1_1+5I^1_2+12I^1_3+8I^1_4 &= A_{5,1},  \tag{5}\\
		C^2_{63}+I^1_1+6I^1_2+10I^1_3+8I^1_4 &= A_{6,1},  \tag{6}\\
		I^1_1+7I^1_2+10I^1_3+5I^1_4 &= A_{7,1},  \tag{7}\\
		I^1_1+8I^1_2+12I^1_3+5I^1_4 &= A_{8,1},  \tag{8}\\
		I^1_1+9I^1_2+3I^1_3+I^1_4 &= A_{9,1},  \tag{9}\\
		I^1_1+10I^1_2+9I^1_3+12I^1_4 &= A_{10,1},  \tag{10}
	\end{align*}
	where $I^1_h = \mathcal{M}_h\cdot U_1^T, h=1,2,3,4$, and $U_1$ is the first row of $U$. 
	
	The user can solve for $I^1_1,I^1_2,I^1_3,I^1_4$ from $(7),(8),(9),(10)$ as they form the equation
	$$\begin{bmatrix}
	1 & 7 & 10 & 5 \\
	1 & 8 & 12 & 5 \\
	1 & 9 & 3 & 1 \\
	1 & 10 & 9 & 12
	\end{bmatrix} \cdot \begin{bmatrix}
	I^1_1 \\ I^1_2 \\ I^1_3 \\ I^1_4 
	\end{bmatrix} =  \begin{bmatrix}
	A_{7,1} \\ A_{8,1} \\ A_{9,1} \\ A_{10,1}
	\end{bmatrix}$$
	where the left matrix is the $(4 \times 4)$ submatrix of $\Psi$ which is invertible. Therefore, the user gets $C^1_{11}$, $C^1_{21}$, and $C^1_{31}$ for record 1 and $C^2_{43}, C^2_{53}$, and $C^2_{63}$ for record 2. Similarly, from subquery 2, the user obtains $C^1_{12}, C^1_{22}$, and $C^1_{32}$ for record 1 and $C^2_{44}, C^2_{54}$, and $C^2_{64}$ for record 2. Hence, the user has all the symbols of $X^1$ which are stored in the node $1, 2, 3$ and all the symbols of $X^2$ which are stored in the node $4, 5, 6$. From the property of regenerating codes, the user can reconstruct $X^1$ and $X^2$ as desired.
	\vspace{10pt}
	
	\begin{table}[h!]
		\centering
		\resizebox{\columnwidth}{!}{
			\begin{tabular}{ |c|c|c|c|c|c|c|c|c|c| } 
				\hline
				node 1 & node 2 & node 3 & node 4 & node 5 & node 6 & node 7 & node 8 & node 9 & node 10 \\
				\hline
				\cellcolor{yellow!25}1 & \cellcolor{yellow!25}1 & \cellcolor{yellow!25}1 &  &  &  &  & & & \\ 
				\hline
				\cellcolor{cyan!25}2 & \cellcolor{cyan!25}2 & \cellcolor{cyan!25}2 &  &  &  &  & & & \\ 
				\hline
				& & & \cellcolor{yellow!25}1 & \cellcolor{yellow!25}1 & \cellcolor{yellow!25}1 &  & & & \\ 
				\hline
				& & & \cellcolor{cyan!25}2 & \cellcolor{cyan!25}2 & \cellcolor{cyan!25}2 &  & & & \\ 
				\hline
				& & & & & & & & & \\ 
				\hline
				& & & & & & & & & \\ 
				\hline
			\end{tabular}}
		\captionsetup{font=scriptsize, width=15cm}   
		\caption{Retrieval pattern for a $(10,3,4,2,1,6)$ MSR code. The $m\alpha \times n$ entries correspond to $C^T$ and the entries labelled by the same number, say $d$, are privately retrieved by subquery $d$.}
		\label{table:1}
	\end{table}
\end{example}

\newpage
Now we give the general construction of our MPIR scheme and prove the decodability and privacy. Recall that we use the MSR code with parameters $$(n,k,r,\alpha,\beta,B)=(pk+(2k-2),k,2k-2,k-1,1,k(k-1))$$ over $\mathbb{F}_q$ to store each record $X^1,\dots,X^m$, which means that $$C^i = \Psi \cdot \mathcal{M}^i$$ where $\mathcal{M}^i$ is the message matrix corresponding to $X^i$ as described in Section 2.2, so $$C = \begin{bmatrix}
\Psi\cdot\mathcal{M}^1 & \cdots & \Psi\cdot\mathcal{M}^m
\end{bmatrix}.$$
Suppose that the user wants to retrieve $p$ records $X^{f_1}, X^{f_2}, \dots, X^{f_p}$. In the retrieval step, the user sends a $(\alpha \times m\alpha)$ query matrix $Q^i$, which we can interpret as $\alpha$ subqueries, to each node $i, i=1,\dots,n$. To form the query matrices, the user generates a $(\alpha \times m\alpha)$ random matrix $U = [u_{ij}]$ whose entries are chosen uniformly at a random from $\mathbb{F}_q$. We choose, for $u \in [p], j \in [k]$, $$V^{(j+(u-1)k)_{f_u}} = I_{\alpha}, $$ and for others $V^{s_t}$ which are not defined above, we choose $V^{s_t} = \textbf{0}_{\alpha \times \alpha}.$ As $$E^{f_u} = \left[\begin{array}{c|c|c}
\textbf{0}_{\alpha \times (f_u-1)\alpha} & I_{\alpha \times \alpha} & \textbf{0}_{\alpha \times (m-f_u)\alpha}
\end{array}\right],$$ we have $$V^{(j+(u-1)k)_{f_u}} E^{f_u} = \left[\begin{array}{c|c|c}
\textbf{0}_{\alpha \times (f_u-1)\alpha} & I_{\alpha \times \alpha} & \textbf{0}_{\alpha \times (m-f_u)\alpha}
\end{array}\right].$$ For the rest, we have $V^{s_t}E^{f_t} = \textbf{0}_{\alpha \times m\alpha}.$ The query matrices are $$Q^i = U + V^{i_{f_1}} E^{f_1} + \dots + V^{i_{f_p}} E^{f_p}, i \in [n].$$ Then, each node computes and returns the length-$\alpha$ $A^T_i = Q^i C^T_i$, and we write $$A_i = (A_{i1},A_{i2},\dots,A_{i\alpha}).$$
\newpage
\textit{Decodability:} The following proof will show the decodability of this scheme. We can see that for $i=1,\dots,n$, 
\begin{align*}
C_i &= \Psi_i \cdot \mathcal{M}\\
&= \Psi_i \cdot \begin{bmatrix}
\longdash \text{ } \mathcal{M}_1 \text{ } \longdash\\
\longdash \text{ } \mathcal{M}_2 \text{ }\longdash\\
\vdots \\
\longdash \text{ } \mathcal{M}_{2k-2} \text{ } \longdash\\
\end{bmatrix} \\
&= \sum_{j=1}^{2k-2} \Psi_{ij} \mathcal{M}_j.
\end{align*}
Thus, $$C^T_i =  \sum_{j=1}^{2k-2} \Psi_{ij} \mathcal{M}_j^T.$$
\vspace{5pt} \\
\noindent Consider first subquery 1, we obtain
\begin{alignat*}{2}
C^{f_1}_{1,(f_1-1)\alpha+1}+\sum_{j=1}^{2k-2} \Psi_{1j} I^1_j  &= 
(U_1+e_{(f_1-1)\alpha+1})C^T_{1} &&= A_{11}, \tag{1}\\
&\vdotswithin{=}   &&\vdotswithin{=} \\
C^{f_1}_{k,(f_1-1)\alpha+1}+\sum_{j=1}^{2k-2} \Psi_{k,j} I^1_j &= (U_1+e_{(f_1-1)\alpha+1})C^T_{k} &&= A_{k,1}, \tag{$k$}\\
&\vdotswithin{=}   &&\vdotswithin{=} \\
C^{f_p}_{(p-1)k+1,(f_p-1)\alpha+1}+\sum_{j=1}^{2k-2} \Psi_{(p-1)k+1,j} I^1_j &= (U_1+e_{(f_p-1)\alpha+1})C^T_{(p-1)k+1} &&= A_{(p-1)k+1,1},  \tag{$(p-1)k+1$} \\
&\vdotswithin{=}   &&\vdotswithin{=} \\
C^{f_p}_{pk,(f_p-1)\alpha+1}+\sum_{j=1}^{2k-2} \Psi_{pk,j} I^1_j &= (U_1+e_{(f_p-1)\alpha+1})C^T_{pk} &&= A_{pk,1},  \tag{$pk$} \\
\sum_{j=1}^{2k-2} \Psi_{pk+1,j}, I^1_j  &= U_1 C^T_{pk+1} &&= A_{pk+1,1},  \tag{$pk+1$} \\
&\vdotswithin{=}   &&\vdotswithin{=} \\
\sum_{j=1}^{2k-2} \Psi_{n,j}, I^1_j  &= U_1 C^T_{n} &&= A_{n,1},  \tag{$n$}
\end{alignat*}
where $I^1_h =  U_1 \cdot \mathcal{M}_h^T , h=1,2,\dots,2k-2$, $U_1$ is the first row of $U$, and $e_t$ is the length-$m\alpha$ binary unit vector with 1 at the $t^{th}$ position. 

The user can solve for $I^1_1,\dots,I^1_{2k-2}$ from $(pk+1),\dots,(n)$ as they form the equation
$$\begin{bmatrix}
\longdash \text{ } \Psi_{pk+1} \text{ }\longdash\\
\vdots \\
\longdash \text{ } \Psi_n \text{ } \longdash\\
\end{bmatrix} \cdot \begin{bmatrix}
I^1_1 \\ I^1_2 \\ \vdots \\ I^1_{2k-2}
\end{bmatrix} =  \begin{bmatrix}
A_{pk+1,1} \\ \vdots \\ A_{n,1}
\end{bmatrix}$$
where, since $n=pk+(2k-2)$, the left matrix is $((2k-2) \times (2k-2))$ square submatrix of $\Psi$ which is invertible by the construction. Note that here we in fact make use of the repair property of the code, which requires $r \times r$ submatrices to be invertible. Therefore, the user gets $$C^{f_j}_{(j-1)k+1,(f_j-1)\alpha+1},\dots,C^{f_j}_{jk,(f_j-1)\alpha+1},$$ i.e., all the symbols of record $f_j$ with label 1 in Table 2 for every $j \in [p]$. Combined with responses from subqueries $i=2,\dots,\alpha$, the user has all the symbols of $X^{f_1}, \dots, X^{f_p}$ which are stored in the first $pk$ nodes. From the recovery property of the regenerating code, the user can finally reconstruct $X^{f_1},\dots,X^{f_p}$ as desired.
\\
\vspace{-5pt}

\textit{Privacy:} As we construct the query matrices $Q^i$ via the random matrix $U$, $Q^i$ is independent from $f_1,\dots,f_p$ which implies that this scheme achieves perfect privacy.

\vspace{10pt}

\begin{table}[h!]
	\centering
	\resizebox{\columnwidth}{!}{
		\begin{tabular}{ |c|c|c|c|c|c|c|c|c| } 
			\hline
			node 1 & $\cdots$ & node $(j-1)k$ & node $(j-1)k+1$ & $\cdots$ & node $jk$ & node $jk+1$ & $\cdots$ & node $n$ \\
			\hline
			& &    & \cellcolor{yellow!25}1 & $\cdots$ & \cellcolor{yellow!25}1 &  &  &  \\ 
			\hline
			& & & \cellcolor{cyan!25}2 & $\cdots$ & \cellcolor{cyan!25}2 &  &  &  \\ 
			\hline
			& & & $\vdots$ & $\ddots$ & $\vdots$ & & &  \\
			\hline
			& &  & \cellcolor{magenta!25}$\alpha$ & $\cdots$ & \cellcolor{magenta!25}$\alpha$ &  &  &  \\ 
			\hline
		\end{tabular}
		}
		\captionsetup{font=scriptsize, width=15cm}   
		\caption{Retrieval pattern for a $(pk+(2k-2),k,2k-2,k-1,1,k(k-1))$ MSR code. The $\alpha \times n$ entries correspond to $(C^{f_j})^T$ and the entries labelled by the same number, say $d$, are privately retrieved by subquery $d$.}
		\label{table:2}
	\end{table}

\subsection{An MPIR scheme using a product-matrix MBR code}

As in \cite{matrix}, Rashmi, Shah, Kumar provided the explicit construction of product-matrix MBR codes for any $k \leq r < n$, we can construct an MPIR scheme using this code with parameters $$(n,k,r,\alpha,\beta,\ell)= \bigg(pk+r,k,r,r,1,\frac{k(2r-k+1)}{2} \bigg)$$ over $\mathbb{F}_q$ to store each record $X^1,\dots,X^m$, i.e., $$C^i = \Psi \cdot \mathcal{M}^i$$ where $\mathcal{M}^i$ is the message matrix corresponding to $X^i$ as described in Section 2.3, so $$C = \begin{bmatrix}
\Psi\cdot\mathcal{M}^1 & \cdots & \Psi\cdot\mathcal{M}^m
\end{bmatrix}.$$ The retrieval step of this scheme and the proof of decodability and privacy are similar to the MPIR schemes in Section 5.1.

\section{Discussion}

In this section we analyse the efficiency of our schemes in Sections 5.1 and 5.2.

\subsection{Analysis of the MPIR scheme using MSR codes}

In this scheme, storage overhead is 
\begin{align*}
\frac{n\alpha}{\ell} &=\frac{(pk+(2k-2))(k-1)}{k(k-1)} \\
&= \frac{pk+(2k-2)}{k} \\
&= (p+2) - \frac{2}{k} < p+2,
\end{align*}
and cPoP equals
\begin{align*}
\frac{dn}{p\ell} &=\frac{(k-1)(pk+(2k-2))}{pk(k-1)} \\
&= 1 + \frac{2k-2}{pk}. \\
\end{align*}
Hence, 
$$ cPoP\bigg(\frac{n-r}{k(SO)}\bigg) = \frac{pk+(2k-2)}{pk} \cdot \frac{pk}{k \big( \frac{pk+(2k-2)}{k}\big)} = 1.$$
This means that our scheme achieves the information theoretic limit as it fits the optimal curve in the trade-off derived in Section 4. Also, as we use the MSR codes in our construction which beneficially reduces the repair cost when a node failure occurs in the system, repair ratio in our scheme is $$\frac{r}{\alpha} = \frac{(2k-2)}{k-1} = 2,$$ which is smaller than PIR schemes that use $(n,k)$-MDS codes (for example, in \cite{MDS}) where the repair ratio equals to $k$ if $k>2$.

\subsection{Analysis of the MPIR scheme using MBR codes}

In this scheme, storage overhead is 
$$
\frac{n\alpha}{\ell} =\frac{(pk+r)(r)}{\frac{k(2r-k+1)}{2}} = \frac{2r(pk+r)}{k(2r-k+1)},
$$
and cPoP equals 
$$
\frac{dn}{p\ell} =\frac{r(pk+r)}{p\frac{k(2r-k+1)}{2}} = \frac{2r(pk+r)}{pk(2r-k+1)}.
$$
However, the smallest storage overhead and cPoP occur when $r=k$ which are $$\frac{2k(p+1)}{k+1} \leq 2(p+1)$$ and $$\frac{2k(p+1)}{p(k+1)} \approx 2,$$ respectively.
And we can see that, 
$$ cPoP\bigg(\frac{n-r}{k(SO)}\bigg) = \frac{2r(pk+r)}{pk(2r-k+1)} \cdot \frac{pk}{k \big( \frac{2r(pk+r)}{k(2r-k+1)}\big)} = 1,$$
which implies that this scheme also meets the information theoretic limit as it fits the optimal curve in the trade-off derived in Section 4. Since we use the MBR codes in our construction, repair ratio in our scheme is $$\frac{r}{\alpha} = 1,$$ which is the smallest possible repair ratio.

\vspace{10pt}
\appendix

\section{An alternative MPIR scheme using product-matrix MSR codes}

Here we give an alternative MPIR scheme. We use the product-matrix MSR codes to store the database in the same way as in Section 5.1 with fewer nodes and different retrieval pattern resulting in lower storage overhead and higher cPoP. However, this scheme also reaches the information theoretic limit in the different point of the optimal curve in the trade-off in Section 4, and it turns out to be a generalisation of our PIR scheme in \cite{MSR}. We first start with an example to illustrate our scheme.

\begin{example}
	Suppose that we have 3 records over the finite field $\mathbb{F}_{13}$, each with size $6$, which can be written as $$X^i = \{x_{i1}, x_{i2}, x_{i3}, x_{i4}, x_{i5}, x_{i6}\}, \mbox{ for } i=1,2,3.$$
	We use a $(8, 3, 4, 2, 1, 6)$ product-matrix MSR code over $\mathbb{F}_{13}$ to encode each record by choosing the encoding matrix $\Psi$ to be the Vandermonde matrix, and the message matrix $\mathcal{M}^i$ for the record $i , i \in \{1,2,3\}$ as described in Section 2.2:
	\[ \Psi =
	\begin{bmatrix}
	1 & 1 & 1 & 1 & 1 & 1 & 1 & 1\\
	1 & 2 & 3 & 4 & 5 & 6 & 7 & 8 \\
	1 & 4 & 9 & 3 & 12 & 10 & 10 & 12 \\
	1 & 8 & 1 & 12 & 8 & 8 & 5 & 5
	\end{bmatrix}^T, \quad  \mathcal{M}^i = \begin{bmatrix}
	x_{i1} & x_{i2} \\
	x_{i2} & x_{i3} \\
	x_{i4} & x_{i5} \\
	x_{i5} & x_{i6} 
	\end{bmatrix}.
	\]
	Hence, each node stores
	\begin{center}
		\scalebox{0.8}{
			\begin{tabular}{ |c|c|c|c| } 
				\hline
				node 1 & node 2 & node 3 & node 4 \\
				\hline
				$x_{11}+x_{12}+x_{14}+x_{15}$ & $x_{11}+2x_{12}+4x_{14}+8x_{15}$ & $x_{11}+3x_{12}+9x_{14}+x_{15}$ & $x_{11}+4x_{12}+3x_{14}+12x_{15}$ \\ 
				$x_{12}+x_{13}+x_{15}+x_{16}$ & $x_{12}+2x_{13}+4x_{15}+8x_{16}$ & $x_{12}+3x_{13}+9x_{15}+x_{16}$ & $x_{12}+4x_{13}+3x_{15}+12x_{16}$ \\ 
				\hline
				$x_{21}+x_{22}+x_{24}+x_{25}$ & $x_{21}+2x_{22}+4x_{24}+8x_{25}$ & $x_{21}+3x_{22}+9x_{24}+x_{25}$ & $x_{21}+4x_{22}+3x_{24}+12x_{25}$ \\ 
				$x_{22}+x_{23}+x_{25}+x_{26}$ & $x_{22}+2x_{23}+4x_{25}+8x_{26}$ & $x_{22}+3x_{23}+9x_{25}+x_{26}$ & $x_{22}+4x_{23}+3x_{25}+12x_{26}$ \\ 
				\hline
				$x_{31}+x_{32}+x_{34}+x_{35}$ & $x_{31}+2x_{32}+4x_{34}+8x_{35}$ & $x_{31}+3x_{32}+9x_{34}+x_{35}$ & $x_{31}+4x_{32}+3x_{34}+12x_{35}$ \\ 
				$x_{32}+x_{33}+x_{35}+x_{36}$ & $x_{32}+2x_{33}+4x_{35}+8x_{36}$ & $x_{32}+3x_{33}+9x_{35}+x_{36}$ & $x_{32}+4x_{33}+3x_{35}+12x_{36}$ \\ 
				\hline
			\end{tabular}}
		\end{center}
		\begin{center}
			\scalebox{0.8}{
				\begin{tabular}{ |c|c|c|c| } 
					\hline
					node 5 & node 6 & node 7 & node 8\\
					\hline
					$x_{11}+5x_{12}+12x_{14}+8x_{15}$ & $x_{11}+6x_{12}+10x_{14}+8x_{15}$ & $x_{11}+7x_{12}+10x_{14}+5x_{15}$ & $x_{11}+8x_{12}+12x_{14}+5x_{15}$\\ 
					$x_{12}+5x_{13}+12x_{15}+8x_{16}$ & $x_{12}+6x_{13}+10x_{15}+8x_{16}$ & $x_{12}+7x_{13}+10x_{15}+5x_{16}$ & $x_{12}+8x_{13}+12x_{15}+5x_{16}$\\ 
					\hline
					$x_{21}+5x_{22}+12x_{24}+8x_{25}$ & $x_{21}+6x_{22}+10x_{24}+8x_{25}$ & $x_{21}+7x_{22}+10x_{24}+5x_{25}$ & $x_{21}+8x_{22}+12x_{24}+5x_{25}$\\ 
					$x_{22}+5x_{23}+12x_{25}+8x_{26}$ & $x_{22}+6x_{23}+10x_{25}+8x_{26}$ & $x_{22}+7x_{23}+10x_{25}+5x_{26}$ & $x_{22}+8x_{23}+12x_{25}+5x_{26}$\\ 
					\hline
					$x_{31}+5x_{32}+12x_{34}+8x_{35}$ & $x_{31}+6x_{32}+10x_{34}+8x_{35}$ & $x_{31}+7x_{32}+10x_{34}+5x_{35}$ & $x_{31}+8x_{32}+12x_{34}+5x_{35}$\\ 
					$x_{32}+5x_{33}+12x_{35}+8x_{36}$ & $x_{32}+6x_{33}+10x_{35}+8x_{36}$ & $x_{32}+7x_{33}+10x_{35}+5x_{36}$ & $x_{32}+8x_{33}+12x_{35}+5x_{36}$\\ 
					\hline
				\end{tabular}}
			\end{center} 
			Recall that $C^a_{ij}$ is the $j^{th}$ symbol of record $a$, stored in node $i$. Here the entire database can be recovered from the content of any 3 nodes, and if any one node failed, it can be repaired by downloading one symbol each from 4 of the remaining nodes.
			\\
			\vspace{-5pt}
			
			In the retrieval step, suppose the user wants record $X^1$ and $X^2$. The query $Q^i$ is a $(3 \times 6)$ matrix which we can interpret as $3$ subqueries submitted to node $i$ for each $i \in [8]$. To form the query matrices, the user generates a $(3 \times 6)$ random matrix $U = [u_{ij}]$ whose entries are chosen uniformly at a random from $\mathbb{F}_{13}$. Choose  \[
			V^{1_1} = V^{4_2} = \begin{bmatrix}
			1 & 0 \\
			0 & 1 \\
			0 & 0  
			\end{bmatrix}, \quad V^{2_1} = V^{5_2} = \begin{bmatrix}
			0 & 0 \\
			1 & 0 \\
			0 & 1 
			\end{bmatrix}, \quad V^{3_1} = V^{6_2} = \begin{bmatrix}
			0 & 1 \\
			0 & 0 \\
			1 & 0
			\end{bmatrix},\]
			\[
			V^{1_2} = V^{2_2} = V^{3_2} = V^{4_1} = V^{5_1} = V^{6_1} = V^{7_1} = V^{7_2} = V^{8_1} = V^{8_2} = \textbf{0}_{3 \times 2}.
			\]
			As \[
			E^1 = \begin{bmatrix}
			1 & 0 & 0 & 0 & 0 & 0 \\
			0 & 1 & 0 & 0 & 0 & 0 \\  
			\end{bmatrix}, \quad E^2 = \begin{bmatrix}
			0 & 0 & 1 & 0 & 0 & 0 \\
			0 & 0 & 0 & 1 & 0 & 0 \\  
			\end{bmatrix},
			\]
			we have
			\[
			V^{1_1}E^1 = \begin{bmatrix}
			1 & 0 & 0 & 0 & 0 & 0 \\
			0 & 1 & 0 & 0 & 0 & 0 \\
			0 & 0 & 0 & 0 & 0 & 0 
			\end{bmatrix}, V^{2_1}E^1 = \begin{bmatrix}
			0 & 0 & 0 & 0 & 0 & 0 \\
			1 & 0 & 0 & 0 & 0 & 0 \\
			0 & 1 & 0 & 0 & 0 & 0
			\end{bmatrix}, V^{3_1}E^1 = \begin{bmatrix}
			0 & 1 & 0 & 0 & 0 & 0 \\
			0 & 0 & 0 & 0 & 0 & 0 \\
			1 & 0 & 0 & 0 & 0 & 0
			\end{bmatrix},
			\]
			\[
			V^{4_2}E^2 = \begin{bmatrix}
			0 & 0 & 1 & 0 & 0 & 0 \\
			0 & 0 & 0 & 1 & 0 & 0 \\
			0 & 0 & 0 & 0 & 0 & 0 
			\end{bmatrix}, V^{5_2}E^2 = \begin{bmatrix}
			0 & 0 & 0 & 0 & 0 & 0 \\
			0 & 0 & 1 & 0 & 0 & 0 \\
			0 & 0 & 0 & 1 & 0 & 0
			\end{bmatrix}, V^{6_2}E^2 = \begin{bmatrix}
			0 & 0 & 0 & 1 & 0 & 0 \\
			0 & 0 & 0 & 0 & 0 & 0 \\
			0 & 0 & 1 & 0 & 0 & 0
			\end{bmatrix},
			\]
			and $$V^{1_2}E^2 = V{2_2}E^2 = V^{3_2}E^2 = V^{4_1}E^1 = V^{5_1}E^1 = $$
			$$V^{6_1}E^1 = V^{7_1}E^1 = V^{7_2}E^2 = V^{8_1}E^1 = V^{8_2}E^2 = \textbf{0}_{3 \times 6}.$$
			
			\noindent The query matrices are $Q^i = U+V^{i_1}E^1+V^{i_2}E^2 , i \in [8]$. Then each node computes and returns the length-$3$ vector $A_i^T = Q^i C^T_i$. Write $A_i = (A_{i1},A_{i2},A_{i3})$. Recall that 
			\vspace{10pt}
			\begin{align*}
			\mathcal{M}_1 &= (x_{11}, x_{12}, x_{21}, x_{22}, x_{31}, x_{32}), \\
			\mathcal{M}_2 &= (x_{12}, x_{13}, x_{22}, x_{23}, x_{32}, x_{33}), \\
			\mathcal{M}_3 &= (x_{14}, x_{15}, x_{24}, x_{25}, x_{34}, x_{35}), \\
			\mathcal{M}_4 &= (x_{15}, x_{16}, x_{25}, x_{26}, x_{35}, x_{36}).
			\end{align*}
			Consider first subquery 1, we obtain
			\begin{align*}
			C^1_{11}+I^1_1+I^1_2+I^1_3+I^1_4 &= A_{11},  \tag{1}\\
			I^1_1+2I^1_2+4I^1_3+8I^1_4 &= A_{21},  \tag{2}\\
			C^1_{32}+I^1_1+3I^1_2+9I^1_3+I^1_4 &= A_{31},  \tag{3}\\
			C^2_{43}+I^1_1+4I^1_2+3I^1_3+12I^1_4 &= A_{41},  \tag{4}\\
			I^1_1+5I^1_2+12I^1_3+8I^1_4 &= A_{51},  \tag{5}\\
			C^2_{64}+I^1_1+6I^1_2+10I^1_3+8I^1_4 &= A_{61},  \tag{6}\\
			I^1_1+7I^1_2+10I^1_3+5I^1_4 &= A_{71},  \tag{7}\\
			I^1_1+8I^1_2+12I^1_3+5I^1_4 &= A_{81},  \tag{8}
			\end{align*}
			where $I^1_h = U_1 \cdot \mathcal{M}_h^T, h=1,2,3,4$, and $U_1$ is the first row of $U$. 
			The user can solve for $I^1_1,I^1_2,I^1_3,I^1_4$ from $(2),(5),(7),(8)$ as they form the equation
			$$\begin{bmatrix}
			1 & 2 & 4 & 8 \\
			1 & 5 & 12 & 8 \\
			1 & 7 & 10 & 5 \\
			1 & 8 & 12 & 5
			\end{bmatrix} \cdot \begin{bmatrix}
			I^1_1 \\ I^1_2 \\ I^1_3 \\ I^1_4 
			\end{bmatrix} =  \begin{bmatrix}
			A_{21} \\ A_{51} \\ A_{71} \\ A_{81}
			\end{bmatrix}$$
			where the left matrix is the $(4 \times 4)$ submatrix of $\Psi$ which is invertible. Therefore, the user gets $C^1_{11}$ and $C^1_{32}$ for record 1 and $C^2_{43}, C^2_{64}$ for record 2. Similarly, from subquery 2, the user obtains $C^1_{12}, C^1_{21}$ for record 1 and $C^2_{44}, C^2_{53}$ for record 2. Lastly, from subquery 3, the user obtains $C^1_{22}, C^1_{31}$ for record 1 and $C^2_{54}, C^2_{63}$ for record 2. Hence, the user has all the symbols of $X^1$ which are stored in the node $1, 2, 3$ and all the symbols of $X^2$ which are stored in the node $4, 5, 6$. From the property of regenerating codes, the user can reconstruct $X^1$ and $X^2$ as desired.
			\vspace{10pt}
			\begin{table}[h!]
				\centering
				\begin{tabular}{ |c|c|c|c|c|c|c|c| } 
					\hline
					node 1 & node 2 & node 3 & node 4 & node 5 & node 6 & node 7 & node 8 \\
					\hline
					\cellcolor{yellow!25}1 & \cellcolor{cyan!25}2 & \cellcolor{magenta!25}3 &  &  &  &  & \\ 
					\hline
					\cellcolor{cyan!25}2 & \cellcolor{magenta!25}3 & \cellcolor{yellow!25}1 &  &  &  &  & \\ 
					\hline
					& & & \cellcolor{yellow!25}1 & \cellcolor{cyan!25}2 & \cellcolor{magenta!25}3 &  & \\ 
					\hline
					& & & \cellcolor{cyan!25}2 & \cellcolor{magenta!25}3 & \cellcolor{yellow!25}1 &  & \\ 
					\hline
					& & &  &  &  &  & \\ 
					\hline
					& & &  &  &  &  & \\ 
					\hline
				\end{tabular}
				\captionsetup{font=scriptsize, width=12.5cm}   
				\caption{Retrieval pattern for a $(8,3,4,2,1,6)$ MSR code. The $6 \times 8$ entries correspond to $C^T$ and the entries labelled by the same number, say $d$, are privately retrieved by subquery $d$.}
				\label{table:3}
			\end{table}
		\end{example}
		
		Next, we will give the general construction of our MPIR scheme and prove the decodability and privacy. In this construction, we use the product-matrix MSR code from \cite{matrix} over the finite field $\mathbb{F}_q$ with $n=(p+2)(k-1)$, i.e. the MSR code with parameters $$(n,k,r,\alpha,\beta,\ell)=((p+2)(k-1),k,2k-2,k-1,1,k(k-1)),$$ to store each record $X^1,\dots,X^m$, which means that $$C^i = \Psi \cdot \mathcal{M}^i$$ where $\mathcal{M}^i$ is the message matrix corresponding to $X^i$ as described in Section 2.2, so $$C = \begin{bmatrix}
		\Psi\cdot\mathcal{M}^1 & \cdots & \Psi\cdot\mathcal{M}^m
		\end{bmatrix}.$$
		Suppose that the user wants to retrieve $p$ records $X^{f_1}, X^{f_2}, \dots, X^{f_p}$ with $p \leq 2k-2$, which we will explain later. In the retrieval step, the user sends a $(k \times m\alpha)$ query matrix $Q^i$, which we can interpret as $k$ subqueries, to each node $i, i=1,\dots,n$. To form the query matrices, the user generates a $(k \times m\alpha)$ random matrix $U = [u_{ij}]$ whose entries are chosen uniformly at a random from $\mathbb{F}_q$. We choose, for $u \in [p]$, $$V^{(1+(u-1)k)_{f_u}} = \left[\begin{array}{c}
		I_{(k-1) \times (k-1)} \\
		\textbf{0}_{1 \times (k-1)}
		\end{array}\right]$$
		
		\vspace{8pt} \noindent
		and $V^{(j+(u-1)k)_{f_u}}, j=2,\dots,k$ is obtained from matrix $V^{((j-1)+(u-1)k)_{f_u}}$ by a single downward cyclic shift of its row vectors. For any $V^{s_t}$ which is not defined above, we choose $V^{s_t} = \textbf{0}_{k \times (k-1)}.$ As $$E^{f_u} = \left[\begin{array}{c|c|c}
		\textbf{0}_{\alpha \times (f_u-1)\alpha} & I_{\alpha \times \alpha} & \textbf{0}_{\alpha \times (m-f_u)\alpha}
		\end{array}\right],$$ we have $$V^{(1+(u-1)k)_{f_u}} E^{f_u} = \left[\begin{array}{cc|c|cc}
		\multicolumn{2}{c|}{\smash{\raisebox{-0.5\normalbaselineskip}{$\textbf{0}_{k \times (f_u-1)\alpha}$}}} & I_{(k-1) \times (k-1)} &         &     \\
		& & \textbf{0}_{1 \times (k-1)} &
		\multicolumn{2}{c}{\smash{\raisebox{.5\normalbaselineskip}{$\textbf{0}_{k \times (m-f_u)\alpha}$}}}
		\end{array}\right]$$
		
		\vspace{8pt} \noindent
		and $V^{(j+(u-1)k)_{f_u}} E^{f_u}, j=2,\dots,k$ is obtained from matrix $V^{((j-1)+(u-1)k)_{f_u}}E^{f_u}$ by a single downward cyclic shift of its row vectors. For the rest, we have $V^{s_t}E^{f_t} = \textbf{0}_{k \times m\alpha}.$ The query matrices are $Q^i = U + V^{i_{f_1}} E^{f_1} + \dots + V^{i_{f_p}} E^{f_p}, i \in [n].$ Then, each node computes and returns the length-$k$ $A^T_i = Q^i C^T_i$, and we write $A_i = (A_{i1},A_{i2},\dots,A_{ik})$.

		\textit{Decodability:} The following proof will show the decodability of this scheme. We can see that for $i=1,\dots,n$, 
		\begin{align*}
		C_i &= \Psi_i \cdot \mathcal{M}\\
		&= \Psi_i \cdot \begin{bmatrix}
		\longdash \text{ } \mathcal{M}_1 \text{ } \longdash\\
		\longdash \text{ } \mathcal{M}_2 \text{ }\longdash\\
		\vdots \\
		\longdash \text{ } \mathcal{M}_{2k-2} \text{ } \longdash\\
		\end{bmatrix} \\
		&= \sum_{j=1}^{2k-2} \Psi_{ij} \mathcal{M}_j
		\end{align*}
		Thus, $$C^T_i =  \sum_{j=1}^{2k-2} \Psi_{ij} \mathcal{M}_j^T.$$
		 Consider first subquery 1, we obtain
		\newpage
		\begin{alignat*}{2}
		C^{f_1}_{1,(f_1-1)\alpha+1}+\sum_{j=1}^{2k-2} \Psi_{1j} I^1_j  &= (U_1+e_{(f_1-1)\alpha+1})C^T_1 &&= A_{1,1},  \tag{1}\\
		\sum_{j=1}^{2k-2} \Psi_{2j} I^1_j  &= U_1 C^T_2 &&= A_{2,1}, \tag{2}\\
		C^{f_1}_{3,(f_1-1)\alpha+k-1}+\sum_{j=1}^{2k-2} \Psi_{3j} I^1_j  &= (U_1+e_{(f_1-1)\alpha+k-1})C^T_3 &&= A_{3,1},  \tag{3}\\
		C^{f_1}_{4,(f_1-1)\alpha+k-2}+\sum_{j=1}^{2k-2} \Psi_{4j} I^1_j &= (U_1+e_{(f_1-1)\alpha+k-2})C^T_4 &&= A_{4,1},  \tag{4}\\
		&\vdotswithin{=}   &&\vdotswithin{=} \\
		C^{f_1}_{k,(f_1-1)\alpha+2}+\sum_{j=1}^{2k-2} \Psi_{k,j} I^1_j &= (U_1+e_{(f_1-1)\alpha+2})C^T_{k} &&= A_{k,1}, \tag{$k$}\\
				&\vdotswithin{=}   &&\vdotswithin{=} \\
		C^{f_p}_{(p-1)k+1,(f_p-1)\alpha+1}+\sum_{j=1}^{2k-2} \Psi_{(p-1)k+1,j} I^1_j  &= (U_1+e_{(f_p-1)\alpha+1})C^T_{(p-1)k+1} &&= A_{(p-1)k+1,1},  \tag{$(p-1)k+1$}\\
		\sum_{j=1}^{2k-2} \Psi_{(p-1)k+2,j} I^1_j  &= U_1 C^T_{(p-1)k+2} &&= A_{(p-1)k+2,1}, \tag{$(p-1)k+2$}\\
		C^{f_p}_{(p-1)k+3,(f_p-1)\alpha+k-1}+\sum_{j=1}^{2k-2} \Psi_{(p-1)k+3,j} I^1_j  &= (U_1+e_{(f_p-1)\alpha+k-1})C^T_{(p-1)k+3} &&= A_{(p-1)k+3,1},  \tag{$(p-1)k+3$}\\
		C^{f_p}_{(p-1)k+4,(f_p-1)\alpha+k-2}+\sum_{j=1}^{2k-2} \Psi_{(p-1)k+4, j} I^1_j &= (U_1+e_{(f_p-1)\alpha+k-2})C^T_{(p-1)k+4} &&= A_{(p-1)k+4, 1},  \tag{$(p-1)k+4$}\\
		&\vdotswithin{=}   &&\vdotswithin{=} \\
		C^{f_p}_{pk,(f_p-1)\alpha+2}+\sum_{j=1}^{2k-2} \Psi_{pk,j} I^1_j &= (U_1+e_{(f_p-1)\alpha+2})C^T_{pk} &&= A_{pk,1}, \tag{$pk$}\\
		\sum_{j=1}^{2k-2} \Psi_{pk+1,j} I^1_j &= U_1 C^T_{pk+1} &&= A_{pk+1,1},  \tag{$pk+1$} \\
		&\vdotswithin{=}   &&\vdotswithin{=} \\
		\sum_{j=1}^{2k-2} \Psi_{n,j}, I^1_j  &= U_1 C^T_{n} &&= A_{n,1},  \tag{$n$}
		\end{alignat*}
		where $I^1_h = \mathcal{M}_h \cdot U_1^T, h=1,2,\dots,2k-2$, $U_1$ is the first row of $U$, and $e_t$ is the length-$m\alpha$ binary unit vector with 1 at the $t^{th}$ position. 
		
		It can be seen that for the first $pk$ nodes, we obtain $p$ linear independent equations from node $2,k+2,\dots,(p-1)k+2$ to use for getting rid of the interferences $I^1_1,\dots,I^1_{2k-2}$. As we want to design our scheme to be optimal, we do not want excessive equations for this job which is the reason why $p$ should not be greater that $2k-2$.\footnote{The scheme still works for the case $p > 2k-2$ but it is not optimal in terms of the trade-off derived in Section 4.} Therefore, the user can solve for $I^1_1,\dots,I^1_{2k-2}$ from $(2),(k+2),\dots,((p-1)k+2),(pk+1),\dots,(n)$ as they form the equation
		$$\begin{bmatrix}
		\longdash \text{ } \Psi_2 \text{ } \longdash\\
				\longdash \text{ } \Psi_{k+2} \text{ } \longdash\\
				\vdots \\
		\longdash \text{ } \Psi_{(p-1)k+2} \text{ }\longdash\\
		\longdash \text{ } \Psi_{pk+1} \text{ }\longdash\\
		\vdots \\
		\longdash \text{ } \Psi_n \text{ } \longdash\\
		\end{bmatrix} \cdot \begin{bmatrix}
		I^1_1 \\ I^1_2 \\ \vdots \\ I^1_{2k-2}
		\end{bmatrix} =  \begin{bmatrix}
		A_{21} \\ A_{k+2,1} \\ \vdots \\ A_{(p-1)k+2,1} \\ A_{pk+1,1} \\ \vdots \\ A_{n,1}
		\end{bmatrix}$$
		where, since $n=(p+2)(k-1)$ and $p \leq 2k-2$, the left matrix is $((2k-2) \times (2k-2))$ square submatrix of $\Psi$ which is invertible by the construction. Note that here we in fact make use of the repair property of the code, which requires $r \times r$ submatrices to be invertible. Therefore, the user gets $$C^{f_j}_{(j-1)k+1,(f_j-1)\alpha+1},C^{f_j}_{(j-1)k+3,(f_j-1)\alpha+k-1}, C^{f_j}_{(j-1)k+4,(f_j-1)\alpha+k-2}, \dots, C^{f_j}_{jk,(f_j-1)\alpha+2},$$ i.e., all the symbols of record $f_j$ with label 1 in Table 4 for every $j \in [p]$. Combined with responses from subqueries $i=2,\dots,k$, the user has all the symbols of $X^{f_1}, \dots, X^{f_p}$ which are stored in the first $pk$ nodes. From the recovery property of the regenerating code, the user can finally reconstruct $X^{f_1}, \dots, X^{f_p}$ as desired.
		\\
		\vspace{-5pt}
		
		\textit{Privacy:} As we construct the query matrices $Q^i$ via the random matrix $U$, $Q^i$ is independent from $f_1,\dots,f_p$ which implies that this scheme achieves perfect privacy.

		\vspace{10pt}
		
		\textit{Analysis:} In this scheme, storage overhead is 
		\begin{align*}
		\frac{n\alpha}{\ell} &=\frac{(p+2)(2k-2)(k-1)}{k(k-1)} \\
		&= \frac{(p+2)(k-1)}{k}  < p+2,
		\end{align*}
		and cPoP equals
		\begin{align*}
		\frac{dn}{p\ell} &=\frac{k(p+2)(k-1)}{pk(k-1)} \\
		&= \frac{p+2}{p} \\
		&= 1 + \frac{2}{p}. \\
		\end{align*}
		Hence, 
		\begin{align*} 
		cPoP\bigg(\frac{n-r}{k(SO)}\bigg) &= \frac{p+2}{p} \cdot \frac{(p+2)(k-1)-(2k-2)}{k \big( \frac{(p+2)(k-1)}{k}\big)} \\
		&= \frac{p+2}{p} \cdot \frac{p(k-1)}{(p+2)(k-1)} \\
		&= 1
		\end{align*}
		which implies that our scheme reaches the information theoretic limit as it fits the optimal curve in the trade-off in Section 4.
		
				\vspace{10pt}
		
		\begin{table}[h!]
			\centering
			\resizebox{\columnwidth}{!}{
				\begin{tabular}{ |c|c|c|c|c|c|c|c|c|c|c|c| } 
					\hline
					node 1 & $\cdots$ & node $(j-1)k$ & node $(j-1)k+1$ & node $(j-1)k+2$ & node $(j-1)k+3$ & $\cdots$ & node $jk-1$ & node $jk$ & node $jk+1$ & $\cdots$ & node $n$ \\
					\hline
					& & & \cellcolor{yellow!25}1 & \cellcolor{cyan!25}2 & \cellcolor{magenta!25}3 & $\cdots$ & \cellcolor{blue!25}$k-1$ & \cellcolor{black!25}$k$ &  & &  \\ 
					\hline
					& & & \cellcolor{cyan!25}2 & \cellcolor{magenta!25}3 & \cellcolor{green!25}4 & $\cdots$ & \cellcolor{black!25}$k$ & \cellcolor{yellow!25}1 &  & & \\ 
					\hline
					& & & $\vdots$ & $\vdots$ & $\vdots$ & $\vdots$ & $\vdots$ & $\vdots$ & & & \\
					\hline
					& & & \cellcolor{red!25}$k-2$ & \cellcolor{blue!25}$k-1$ & \cellcolor{black!25}$k$ & $\cdots$ & \cellcolor{blue!50}$k-4$ & \cellcolor{magenta!50}$k-3$ & & & \\
					\hline
					& & & \cellcolor{blue!25}$k-1$ & \cellcolor{black!25}$k$ & \cellcolor{yellow!25}1 & $\cdots$ & \cellcolor{magenta!50}$k-3$ & \cellcolor{red!25}$k-2$ &  & & \\
					\hline
				\end{tabular}}
				\captionsetup{font=scriptsize, width=15cm}   
				\caption{Retrieval pattern for a $((p+2)(k-1),k,2k-2,k-1,1,k(k-1))$ MSR code. The $\alpha \times n$ entries correspond to $(C^{f_j})^T$ and the entries labelled by the same number, say $d$, are privately retrieved by subquery $d$. \vspace{10pt}}
				\label{table:4}
			\end{table}

	We lastly note that the retrieval technique in this paper can also be applied to the single-PIR scheme using MDS codes in \cite{MDS} to retrieve multiple records.

\end{document}